\documentclass[aps,superscriptaddress,twocolumn,floatfix,letterpaper,nofootinbib]{revtex4-1}

\usepackage{braket}

\newcommand{\RZ}{{\mathbb{R}/\mathbb{Z}}}

\newcommand\hcup[1]{\underset{{\scriptscriptstyle #1}}{\smile}}
\newcommand\toZ[1]{\lfloor #1 \rceil}

\newcommand\beq{\begin{equation}}
\newcommand\eeq{\end{equation}}

\usepackage[dvipsnames]{xcolor}

\begin{document}

\begin{titlepage}

\title{
A Commuting Projector Model with a Non-zero Quantized Hall conductance
}

\author{Michael DeMarco}
\affiliation{Department of Physics, Massachusetts Institute of
Technology, Cambridge, MA 02139, USA}
\email{demarco@mit.edu}

\author{Xiao-Gang Wen}
\affiliation{Department of Physics, Massachusetts Institute of
Technology, Cambridge, MA 02139, USA}
\email{xgwen@mit.edu}

\begin{abstract} 

By ungauging a recently discovered lattice rotor model for Chern-Simons theory,
we create an exactly soluble path integral on spacetime lattice for
$U^\kappa(1)$ Symmetry Protected Topological (SPT) phases in $2+1$ dimensions
with a non-zero Hall conductance.  We then convert the path integral on a
$2+1$d spacetime lattice into a $2$d Hamiltonian lattice model, and show that
the Hamiltonian consists of mutually commuting local projectors.  We confirm
the non-zero Hall conductance by calculating the Chern number of the exact
ground state.  It has recently been suggested that no commuting projector model
can host a nonzero Hall conductance. We evade this no-go theorem by considering a rotor model, with a countably infinite number of states per site.

\end{abstract}


\maketitle

\end{titlepage}

\noindent 

The discovery of the Quantum Hall Effect ignited a revolution in condensed
matter physics \cite{Klitzing:2020ab, RevModPhys.71.S298, 10.1007/BFb0113369}.
Both the integral and fractional \cite{PhysRevLett.48.1559} forms showed the
limits of the symmetry-breaking approach to phases of matter
\cite{W8987,W9039,2004cond.mat.11275S} and new research revealed states of
matter hosting fractional particles \cite{PhysRevLett.50.1395, STERN2008204},
protected edge modes \cite{W9505}, and topological
ground state degeneracy \cite{W8987}. Together with research directed at
understanding high-$T_c$ superconductors \cite{WWZ8913,RS9173,W9164,
PhysRevB.62.7850}, the new era of condensed matter has
unveiled physics which is stranger, richer, and more entangled than before.
Certainly, more is indeed different. 

Commuting projector models have been a central tool for understanding the new
zoo of theories. Employed most famously by Kitaev \cite{Kitaev:1997wr} to
provide an exactly soluble model for the previously proposed 2+1d $Z_2$
topological order \cite{RS9173,W9164} with emergent fermions and anyons, they
now describe models for a wide class of string-net topological order
\cite{Levin:2004mi}, recently unleashed a flurry of research on fractons
\cite{Cc0404182,Vijay:2015mka}, and continue to underlie our microscopic
understanding of exotic phases.

It is quite surprising then that no commuting projector model has been
discovered for gapped phases with non-zero Hall conductance. It was commonly
believed that none could exist, and recently a no-go theorem has been proposed
\cite{Kapustin:2020aa}, ruling out a large class of potential theories with a
finite Hilbert space on each site. 

In this paper, we describe a commuting projector model for $U^\ka(1)$ SPT
phases with non-zero quantized Hall conductance
\cite{CGL1314,LV1219,LW1305,SL1301}, providing an exactly solved model for
$U^\ka(1)$ SPT states. Related to the evasion of the no-go theorem, the
physical degrees of freedom in our model are $U^\ka(1)$ rotors, with a
countably infinite Hilbert space on each site. There is no clear way to reduce
the on-site Hilbert space to finite dimension, while retaining $U^\ka(1)$
symmetry and commuting projector property, because the Lagrangian or the
Hamiltonian, while local, is not a smooth function of the rotor variables,  

The phases we describe are short-range-entangled SPT phases, with a unique
ground state on any manifold. Employing the recent discovery of a rotor path
integral \cite{PhysRevLett.126.021603} for $K$-matrix Chern-Simons theory, we
ungauge \cite{W181202517} that model to derive another rotor path integral for
general $U^\ka(1)$ SPT phases. In the Hamiltonian approach, the latter rotor
path integral yields an exactly soluble  commuting projector lattice model. 

First, we review the Chern-Simons model and its properties. We then ungauge
\cite{W181202517} it in the presence of a background field and show that
integrating out the matter fields leads to a Chern-Simons response term. A
conversion from the Lagrangian to Hamiltonian approaches then yields a
commuting projector Hamiltonian for the most general $U^\ka(1)$ SPT state. The
Hamiltonian model leads to a unique wavefunction, which we show has the
expected Chern number over the ``holonomy torus.'' Finally, we discuss the
relation of our model to the no-go theorem and the theory of discontinuous
group cocycles.

\emph{Chern-Simons Lattice Model.} In \Ref{PhysRevLett.126.021603}, a local
bosonic rotor model was constructed which realizes a $2+1$d topological order
described by $U^\ka(1)$ Chern-Simons theory with an even $K$-matrix. The
topological order has a chiral central charge given by the signature of the
$K$-matrix, and hosts an exact $Z_{k_1}\times Z_{k_2} \times \cdots$
1-symmetry, where $k_i$ are the diagonal entries of the Smith normal form of
$K$.

The model is formulated in terms of cochains on a spacetime simplicial complex.
A spacetime complex (lattice) is a triangulation of the three-dimensional
space-time \emph{with a branching structure},
\cite{C0527,CGL1314}, denoted as $\cM^3$. We denote
vertices of the complex as $i,j,\cdots$, links as $ \<ij\>$, and so forth. For
the Chern-Simons Lattice model with a $\ka\times\ka$ $K$-matrix, the physical
degrees are rotor one-cochains, i.e. $a_{I, ij}^\RZ$ on each link $\<ij\>$
(which corresponds to latice gauge field), $I = 1...\kappa$. For the ungauged
model they will be rotor zero-cochains, i.e. a $\phi_{I, i}^\RZ$ on each site
(which corresponds to latice scaler field). We work in units where flux and cycles are quantized to unity; for instance, $U^\kappa(1)$ variables may be obtained as $u_{I, ij}^{U^\kappa(1)} = e^{2\pi i a_{I, ij}^\RZ}$, $\varphi_{I, ij}^{U^\kappa(1)} = e^{2\pi i \phi_{I, ij}^\RZ}$. Accordingly, we will require that
all quantities be invariant under a gauge redundancy:
\begin{align}
a^\RZ_{I, ij} \to a^\RZ_{I, ij} + n_{I, ij}, \ \ \ \ 
\phi^\RZ_{I, i} \to \phi^\RZ_{I, i} + n_{I, i}
\label{Zredundancy}
\end{align}
with $n_{I, ij}, n_{I, i}\in \mathbb Z$, so that these variables are genuinely
$\RZ$-valued.  In this case, the lattice gauge fields $a^\RZ_{I, ij}$ will
describe a compact $U^\ka(1)$ gauge theory, and the  lattice scaler fields
$\phi^\RZ_{I, i}$ will describe a bosonic model with $U^\ka(1)$ symmetry.

We will make use of the lattice differential $\dd$ \cite{Hatcher:478079} which
sends $m$-cochains $f_m$ to $m+1$ cochains $\dd f_m$ and satisfies $\dd^2=0$,
and the lattice cup k-products, which take an $m$-cochain $f_m$ and a
$n$-cochain $g_n$ and return a $n+m - k$ cochain $f_m\cup_k g_n$. We will
abbreviate by omitting zero-cup products $f_m\cup_0 g_n = f_m g_n$. For a
review of this notation, see \cite{Hatcher:478079} and the supplemental
material of  \cite{PhysRevLett.126.021603}.

We can now write down the Cher-Simons lattice model.  Given a bosonic
$K$-matrix $K_{IJ}$ with even diagonal entries, define 
$k_{IJ} = K_{IJ} \text{ for } I\neq J$ and
$k_{IJ} = \frac{1}{2}K_{IJ} \text{ for } I = J$.
In terms of this reduced matrix $k_{IJ}$, the path integral is:
\begin{align}
\label{CSlatt}
& Z =\int [\prod\dd a^{\RZ}_I]\ 
\ee^{\ii 2\pi \sum_{I\leq J} k_{IJ} \int_{\cM^3} \dd \big(a^{\RZ}_I(a^{\RZ}_J-\toZ{a^{\RZ}_J}  )\big) }
\nonumber\\
&
 \ee^{\ii 2\pi \sum_{I\leq J} k_{IJ} \int_{\cM^3} a^{\RZ}_{I} (\dd a^{\RZ}_{J} -\toZ{\dd a^{\RZ}_J})-\toZ{\dd a^{\RZ}_I}a^{\RZ}_J }
\\
&
\ee^{-\ii 2\pi \sum_{I\leq J} k_{IJ} \int_{\cM^3} a^{\RZ}_J\hcup{1}\dd \toZ{\dd a^{\RZ}_I}}  
\ee^{- \int_{\cM^3} \frac{|\dd a^{\RZ}_I - \toZ{\dd a^{\RZ}_I}|^2}{g}}
\nonumber 
\end{align}
Here $\toZ{x}$ denotes the nearest integer to $x$ and $\int_{\cM^3}$ means the signed
sum over all $3$-simplices in $\cM^3$ induced by evaluation against the fundamental class. The measure $\int [\prod\dd a^{\RZ}_I] \equiv
\prod_{\<ij\>}\prod_I \int_{-\frac12}^{\frac12}\dd a^{\RZ}_{I,ij}$ gives rise
to the path integral, where $\prod_{\<ij\>}$ is a product over all the links.
The Maxwell term $\ee^{- \int_{\cM^3} \frac{|\dd a^{\RZ}_I - \toZ{\dd
a^{\RZ}_I}|^2}{g}}$ is included to make $\dd a^{\RZ}_I$ nearly an integer if we
choose $g$ to be small, which realizes the semi-classical limit.  

The invariance of (\ref{CSlatt}) under eq. (\ref{Zredundancy}), its higher symmetry, and its semiclassical limit are described in Ref. \cite{PhysRevLett.126.021603}. Here we focus on the SPT state obtained by ungauging. 

\emph{SPT State from Ungauging.} To ungauge the model of (\ref{CSlatt}), we set
\cite{W181202517}
\begin{align}
 a_{I,ij}^\RZ = \phi_{I,i}^\RZ - \phi_{I,j}^\RZ
,
\ \ \text{or }\ \
a_I^\RZ = \dd \phi^\RZ_I, 
\end{align}
where $\phi^\RZ_I$ are zero-cochains defined on the vertices. 
The partition function is now given by:
\begin{align}
\label{U1spt}
 Z &=\int [\prod\dd \phi^{\RZ}_I]\ 
 \ee^{\ii 2\pi \sum_{I\leq J} k_{IJ} \int_{\cM^3} 
\dd \phi^\RZ_I \dd \toZ{\dd \phi^\RZ_J} 
}
\end{align}
where $\cM^3$ may have boundaries, and the measure is taken to be integration
over all sites, $ \int \prod\dd \phi^{\RZ}_I = \prod_{i} \prod_I
\int_{-\frac{1}{2}}^\frac{1}{2}\dd \phi^\RZ_I$. Eq. (\ref{U1spt}) is the path
integral description of the commuting projector model. As the action is a total
derivative, the partition function is unity on any closed manifold, implying
that the model describes a trivial topological order with zero central charge. On any spatial boundary, the path integral defines a wavefunction $\ket \psi$, where
\begin{equation}
\braket{\phi_{I, i}^\RZ| \psi}
=
\exp\left\{-2\pi i \sum_{I\leq J} k_{IJ} \int_{\cM^2} d\phi_{I}^\RZ \toZ{d\phi_{J}^\RZ}\right\}
\label{eq:gs}
\end{equation}
This will be the ground state of the commuting projector model.

The model has a $U^\kappa(1)$ symmetry:
\begin{equation}
\phi_{I, i}^\RZ \to  \phi_{I, i}^\RZ + \theta_I
\label{eq:Rsymm}
\end{equation}
for constant $\theta_I$.  On a manifold with a boundary, the model is invariant
under eq. (\ref{Zredundancy}) if and only if the reduced $k$-matrix $k_{IJ}$ is
integral, i.e. if the original $K$-matrix $K_{IJ}$ is integral with even
diagonals (which turns out to describe a quantuized Hall conductance and a
bosonic SPT order). In this case, the field
variables are indeed $\RZ$ valued. A reduction of this model to a known one for
$\mathbb Z_n $ SPT states \cite{CGL1314} is given in the
Supplemental material. 

In fact, the model realizes a $U^\kappa(1)$ SPT state. To see the SPT order, we
repeat the ungauging in the presence of a weak background gauge field $\bar
a_{I}^\RZ$ and evaluate the effective action for $\bar a_I^\RZ$. In the
presence of background $U^\ka(1)$ background gauge field $\bar a_I^\RZ$, the
ungauing is done via 
\begin{align} 
a_I^\RZ = \bar a_I^\RZ + \dd \phi^\RZ_I.
\end{align} 
Now the model is given by
\begin{align}
\label{U1sptG}
 Z &=
 \ee^{\ii 2\pi \sum_{I\leq J} k_{IJ} \int_{\cM^3} \bar a^{\RZ}_{I} (\dd \bar a^{\RZ}_{J} -\toZ{\dd \bar a^{\RZ}_J})-\toZ{\dd \bar a^{\RZ}_I}\bar a^{\RZ}_J }
\nonumber \\
&\ \ \ \
\ee^{-\ii 2\pi \sum_{I\leq J} k_{IJ} \int_{\cM^3} \bar a^{\RZ}_J\hcup{1}\dd \toZ{\dd \bar a^{\RZ}_I}}  
\\
&\ \ \ \ 
\int [\prod\dd \phi^{\RZ}_I]\ 
\ee^{-\ii 2\pi \sum_{I\leq J} k_{IJ} \int_{\cM^3} \dd\phi^{\RZ}_J\hcup{1}\dd \toZ{\dd \bar a^{\RZ}_I}} 
\nonumber \\
&\ \ \ \
\ee^{\ii 2\pi \sum_{I\leq J} k_{IJ} \int_{\cM^3} \dd \phi^{\RZ}_{I} (\dd \bar a^{\RZ}_{J} -\toZ{\dd \bar a^{\RZ}_J})-\toZ{\dd \bar a^{\RZ}_I}\dd \phi^{\RZ}_J }
\nonumber\\
&
\ee^{\ii 2\pi \sum_{I\leq J} k_{IJ} \int_{\prt \cM^3} 
(\bar a^{\RZ}_I +\dd \phi_I^\RZ)(\bar a^{\RZ}_J+\dd \phi_J^\RZ-\toZ{\bar a^{\RZ}_J +\dd \phi_J^\RZ}  ) }
\nonumber 
\end{align}
If $\cM^3$ is closed, this can be simplified to
\begin{align}
 Z &=
 \ee^{\ii 2\pi \sum_{I\leq J} k_{IJ} \int_{\cM^3} \bar a^{\RZ}_{I} (\dd \bar a^{\RZ}_{J} -\toZ{\dd \bar a^{\RZ}_J})-\toZ{\dd \bar a^{\RZ}_I}\bar a^{\RZ}_J }
\nonumber \\
&\ \ \ \
\ee^{-\ii 2\pi \sum_{I\leq J} k_{IJ} \int_{\cM^3} \bar a^{\RZ}_J\hcup{1}\dd \toZ{\dd \bar a^{\RZ}_I}}  
\\
&\ \ \ \ 
\int [\prod\dd \phi^{\RZ}_I]\ 
\ee^{-\ii 2\pi \sum_{I\leq J} k_{IJ} \int_{\cM^3} \dd\phi^{\RZ}_J\hcup{1}\dd \toZ{\dd \bar a^{\RZ}_I}} 
\nonumber \\
&\ \ \ \
\ee^{-\ii 2\pi \sum_{I\leq J} k_{IJ} \int_{\cM^3} \dd \phi^{\RZ}_{I} \toZ{\dd \bar a^{\RZ}_J} + \toZ{\dd \bar a^{\RZ}_I}\dd \phi^{\RZ}_J }
\nonumber 
\end{align}
Now we assume that the background field is ``weak.'' Because we are working with $\RZ$ valued fields, a weak field means that $\dd\bar a_I^\RZ$ is nearly an integer, i.e. $|\dd\bar a_I^\RZ - \toZ{\dd\bar a_I^\RZ}| < \epsilon$. Noting that this implies that $\dd\toZ{\dd \bar a_I^\RZ} = 0$, the path integral becomes: 
\begin{align}
\label{U1sptInv}
 Z &=
 \ee^{\ii 2\pi \sum_{I\leq J} k_{IJ} \int_{\cM^3} \bar a^{\RZ}_{I} (\dd \bar a^{\RZ}_{J} -\toZ{\dd \bar a^{\RZ}_J})-\toZ{\dd \bar a^{\RZ}_I}\bar a^{\RZ}_J } 
\end{align}
This is the Chern-Simons response on lattice. When $\cM^3$ is closed, the
action is invariant under gauge transformations of the background gauge field
$\bar a_I^\RZ \to \bar a_I^\RZ + \dd\varphi_I^\RZ$. If $\cM^3$ is a disk, the
redundancy (\ref{Zredundancy}) can be used to set $\toZ{\dd\bar a_I^\RZ} = 0$,
and the response becomes:
\begin{align}
\label{U1sptInv2}
 Z &=
 \ee^{\ii \pi \sum_{I, J} K_{IJ} \int_{\cM^3} \bar a^{\RZ}_{I} \dd \bar a^{\RZ}_{J}}
\end{align}
This Chern-Simons response, in terms of the unreduced $K$-matrix $K_{IJ}$,
describes the Hall conductance and is the SPT invariant for our model. 

Given that the bulk behavior of the path integral is trivial, we expect that we
should be able to create an exactly soluble Hamiltonian model to describe the
time-evolution. Furthermore, because the path integral defines a wavefunction
on any spatial boundary independently of the bulk dynamics, we expect that this
Hamiltonian model should be a commuting projector onto a ground state. As we
shall now see, both are true. 

\begin{figure}[b]
\includegraphics[width = .8\columnwidth]{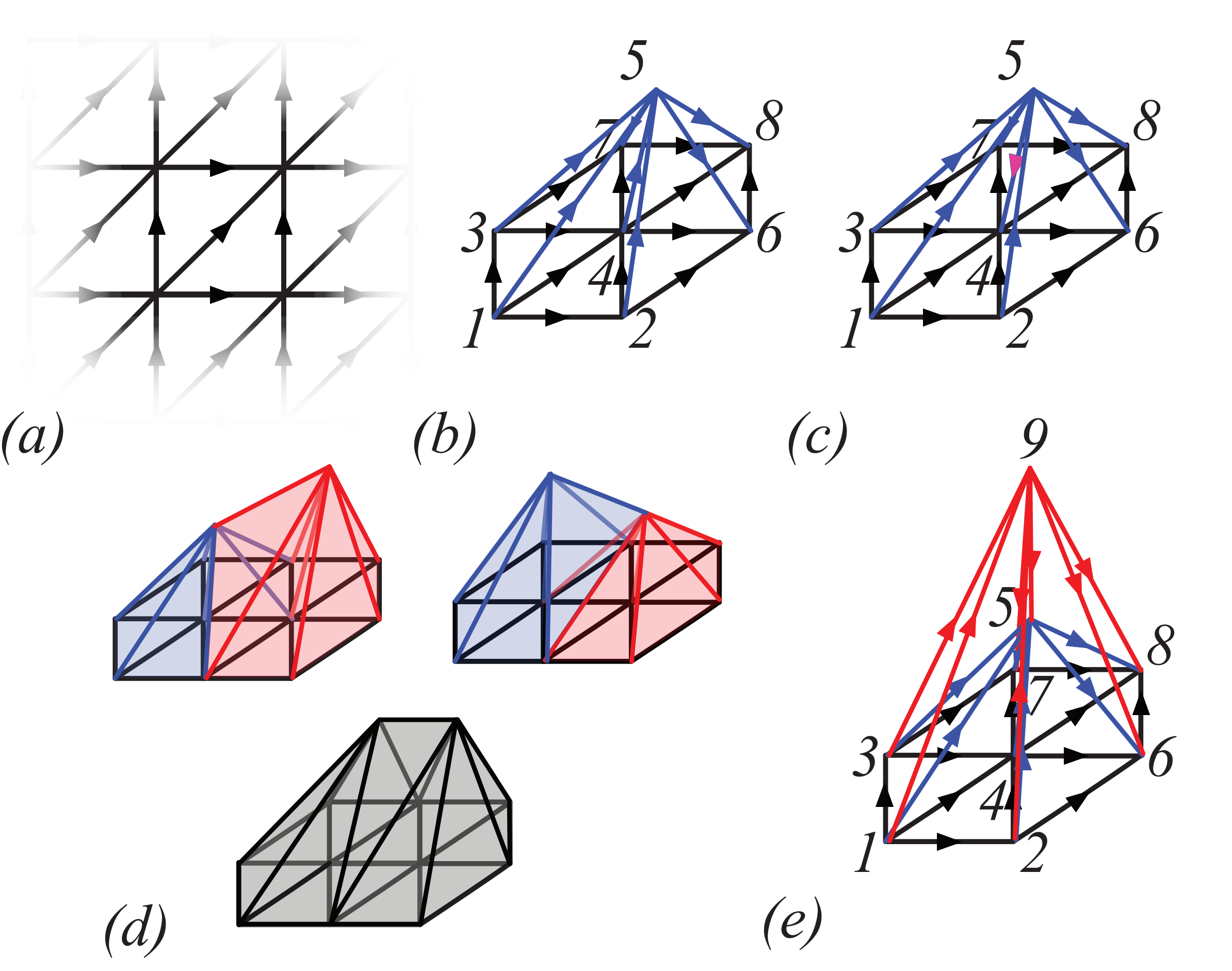}
\caption{\emph{(Color Online).} We construct a commuting projector model for
the lattice in $(a)$ by evaluating the spacetime path integral for the complex
in $(b)$ and turning the amplitude into operators. Because the path integral
contains only a surface term, we can show that $(c)$ the matrix is hermitian,
(d) the operators commute, and $(e)$ they are projectors.}
\label{fig:Hamiltonian_Lattice}
\end{figure}

\emph{Commuting Projector Model.} We can use the spacetime formalism to
construct a commuting projector Hamiltonian on a triangular lattice of the sort
shown in Fig. \ref{fig:Hamiltonian_Lattice}a.  To do so, we consider the time
evolution of a single site, $\phi_4^\RZ \to \phi_5^\RZ$, while preserving the
orientation of lattice links as shown in fig \ref{fig:Hamiltonian_Lattice}b.
Evaluating the path integral on the complex shown yields the matrix elements
for the transition $\phi_4^\RZ \to \phi_5^\RZ$ as a function of the surrounding
$\phi_i^\RZ$:
\begin{widetext}
\begin{multline}
M_{\phi_{I, 4}^\RZ \to \phi_{I, 5}^\RZ}(\phi_{I, 1}^\RZ, ..., \phi_{I, 8}^\RZ)
=
\exp\Big\{
2 \pi i \sum_{I\leq J}k_{IJ}\Big[
  \phi_{I,0}^\RZ\Big(\toZ{\phi_{J,5}^\RZ - \phi_{J,2}^\RZ} + \toZ{\phi_{J,3}^\RZ - \phi_{J,5}^\RZ}  + \toZ{\phi_{J,4}^\RZ-\phi_{J,3}^\RZ} + \toZ{\phi_{J,2}^\RZ - \phi_{J,4}^\RZ}\Big)\\ 
+ \phi_{I,2}^\RZ\Big(\toZ{\phi_{J,5}^\RZ - \phi_{J,6}^\RZ} + \toZ{\phi_{J,2}^\RZ - \phi_{J,5}^\RZ}  + \toZ{\phi_{J,4}^\RZ - \phi_{J,2}^\RZ} + \toZ{\phi_{J,6}^\RZ - \phi_{J,4}^\RZ}\Big)
+ \phi_{I,3}^\RZ\Big(\toZ{\phi_{J,7}^\RZ - \phi_{J,5}^\RZ} + \toZ{\phi_{J,4}^\RZ - \phi_{J,7}^\RZ} \\  + \toZ{\phi_{J,3}^\RZ - \phi_{J,4}^\RZ} + \toZ{\phi_{J,5}^\RZ - \phi_{J,3}^\RZ}\Big)
+ \phi_{I,5}^\RZ\Big(\toZ{\phi_{J,6}^\RZ - \phi_{J,5}^\RZ} + \toZ{\phi_{J,8}^\RZ - \phi_{J,6}^\RZ}   + \toZ{\phi_{J,7}^\RZ - \phi_{J,8}^\RZ} + \toZ{\phi_{J,5}^\RZ - \phi_{J,7}^\RZ}\Big)\\ 
+ \phi_{I,4}^\RZ\Big(\toZ{\phi_{J,7}^\RZ - \phi_{J,4}^\RZ} + \toZ{\phi_{J,8}^\RZ - \phi_{J,7}^\RZ} + \toZ{\phi_{J,6}^\RZ - \phi_{J,8}^\RZ} + \toZ{\phi_{J,4}^\RZ - \phi_{J,6}^\RZ}\Big)
\Big]
\Big\}\label{eq:M_itoj}
\end{multline}
\end{widetext}
We will interpret this transition amplitude as the matrix element for an
operator $\hat M_4$ acting on site-4. However, eq. (\ref{eq:M_itoj}) is
somewhat daunting. Let us set consider $\hat M_4$ as an operator acting only on
the Hilbert space on site-4. If $k_{IJ}=0$, then $\bra{\phi_{I,4}'}\hat
M_4\ket{\phi_{I, 4}} = 1$, and $\hat M_4$ is simply the projector onto the
state with zero angular momentum in each $U(1)$. For nonzero $k_{IJ}$, we may
rewrite the transition amplitude as $M_{\phi_{I, 4}^\RZ \to \phi_{I, 5}^\RZ} =
\exp(2\pi i (f(\phi_{I, 5}^\RZ) -f(\phi_{I, 4}^\RZ))$ (note that this implies hermiticity) where $f(\phi)$ is a
function defined in the appendix that depends on $\phi$ and takes as parameters
$ \phi_{I, 1}^\RZ...\phi_{I, 8}^\RZ$, but not $ \phi_{I, 4}^\RZ$ or $\phi_{I,
5}^\RZ$. This implies that, up to an overall phase, the $\hat M_i$ act as
\begin{equation}
\hat M \ket{\phi_{I}} \propto \int d\phi_{I}e^{2\pi i f(\phi_{I})} \ket{\phi_{I}}
\end{equation}
We see then that these projectors may be thought of as `twisting' the zero
angular momentum state by an phase function $f(\phi_I)$ which depends on the
surrounding values of $\phi_{I, i}$. The phase itself is determined by the
cocycle of the action in (\ref{U1spt}). 

We may construct $\hat M_i$ for an entire lattice. As shown in the supplemental
material, the $\hat M_i$ are hermitian and $U^\ka(1)$ symmetric under $\phi_{I,
i}^\RZ \to \phi_{I, i}^\RZ + \theta_{I} + n_{I, i}$.  

The $\hat M_i$ inherit a number of remarkable properties from the fact that the
$2+1$d path integral action contains only a surface term. First, they mutually
commute: consider the three spacetime complexes in figure
\ref{fig:Hamiltonian_Lattice}c, which addresses the only nontrivial case of two
adjacent $\hat M_i$. The two colored complexes correspond to time evolving
either the blue site followed by the red or the red followed by the blue,
respectively. Because the action contains only a surface term, and the surfaces
are identical, it assigns the same amplitude to both cases. Back in the
Hamiltonian picture, this implies that $[\hat M_i, \hat M_j] = 0$.

For the same reason, the $\hat M_i$ are projectors. The fundamental mechanism
is illustrated in Fig. \ref{fig:Hamiltonian_Lattice}d, where we see the
effect of time-evolving twice. In the language of eq. \ref{eq:M_itoj}, this is
the expression $M_{\phi_{I,5}^{\RZ}\to\phi_{I,
9}^\RZ}M_{\phi_{I,4}^{\RZ}\to\phi_{I, 5}^\RZ}$; in the Hamiltonian picture this
is $\hat M_i^2$. However, because the path integral action depends only on the
values of $\phi_I^\RZ$ on the surface, we could equally well drop site $5$ and
its associated links; the amplitude will not change. This implies that 
$\hat M_i^2 = \hat M_i$. 

The aforementioned hermiticity of the  $\hat M_i$ is also due to this reason, as reversing the orientation
of the link $\<4,5\>$ in Fig. \ref{fig:Hamiltonian_Lattice}c changes the amplitude by complex conjugation. All of these properties of the $\hat M_i$ are demonstrated in the
Supplemental Material.  

Using the mutually commuting projectors $\hat M_i$, we can define a Hamiltonian 
\begin{equation}
H = -g\sum_i \hat M_i
\end{equation}
to obtain a system with ground state $\ket \psi$ from eq. (\ref{eq:gs}) and gap $g$.

With a model producing the ground state eq. (\ref{eq:gs}) now in hand, we return to the Hall
conductance. We have already argued from the path integral that coupling the
system to a background gauge field leads to the expected Chern-Simons response
function, but here we appeal directly to the wavefunction of the system on a
torus to confirm the hall conductance. 

\begin{figure}[t]
\includegraphics[width = .9\columnwidth]{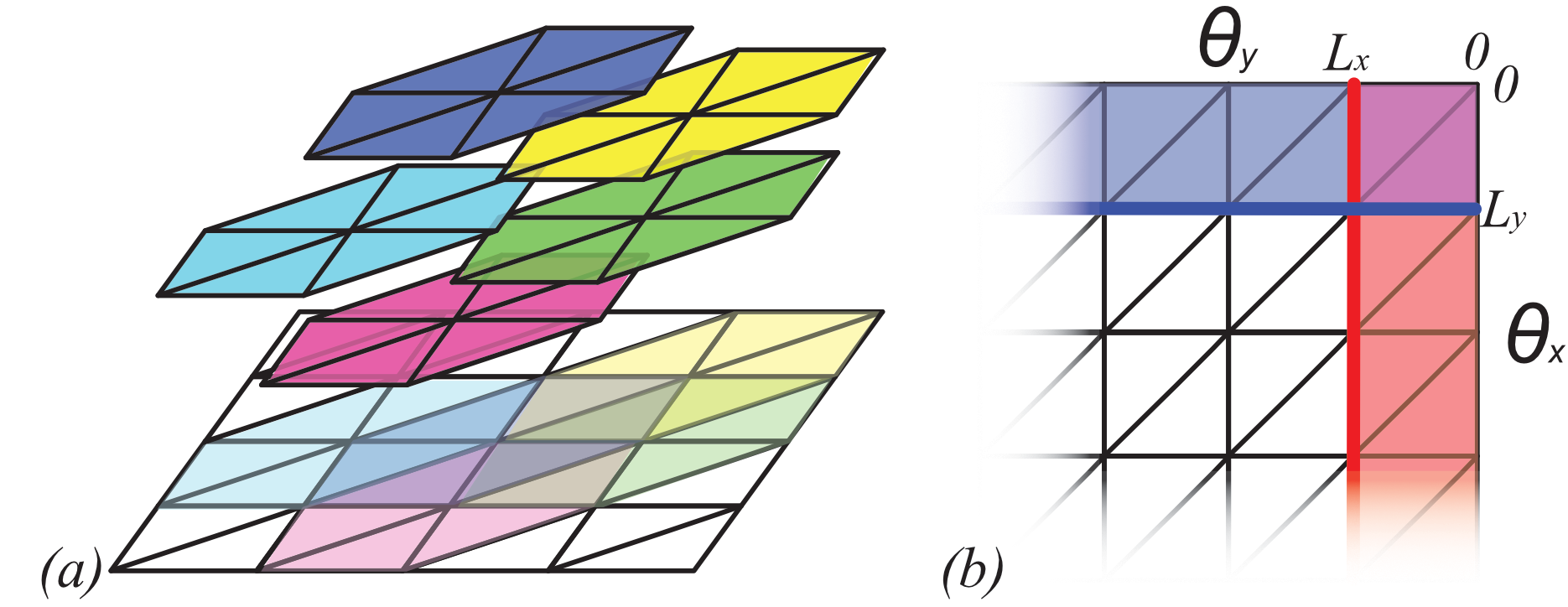}
\caption{\emph{(Color Online)}. $(a)$ The commuting projectors act on a hexagon to define a ground state. $(b)$ To calculate the Chern number of the ground state wavefunction over the holonomy torus, we twist the boundary conditions by $\theta_x n_{I}$ and $\theta_y m_I$ in the $x$ and $y$ directions, respectively.}
\label{fig:HCSU}
\end{figure}

Consider the wavefunction on the lattice shown in Fig. \ref{fig:HCSU}. We twist the boundary conditions by $\theta_x n_I$, $\theta_y
m_I$, 
so that $\phi_{I, x=0, y}^\RZ= \phi_{I, x=L_x, y}^\RZ - \theta_x n_{I}$ and $\phi_{I,x, y=0}^\RZ = \phi_{I, x, y=L_y}^\RZ - \theta_y m^I$,
where the integral vectors $n_I, m_I$ allow us
to describe both direct ($n_I = m_I$) and mixed hall responses. Denote the
ground state with boundary conditions $\theta_x, \theta_y$ by $\ket{\theta_x,
\theta_y}$. 
We are interested in the phase accumulated when we twist the boundary
conditions by an integer, sending $\theta_x \to \theta_x + \ell_x, \theta_y \to
\theta_y + \ell_y$ with $\ell_x, \ell_y \in \mathbb Z$. As shown in the Appendix, the
wavefunction transforms as:
\begin{multline}
\ket{\theta_x+\ell_x, \theta_y+\ell_y} =
\\
e^{-2\pi i \sum_{I \leq J}k_{IJ} \left[
n_I \theta_x m_J\ell_y
-m_I\theta_y n_J\ell_x
\right] }
\ket{\theta_x, \theta_y}
\label{eq:ChernPhase}
\end{multline}
To make sense of this, apply the gauge transformation:
\begin{equation}
\ket{\theta_x, \theta_y} \to e^{-2\pi i \sum_{I\leq J}k_{IJ} n_J m_I \theta_x \theta_y} \ket{\theta_x, \theta_y} \
\end{equation}
In this gauge, one may replace the phase in eq. (\ref{eq:ChernPhase}) by:
\begin{equation}
e^{-2\pi i \sum_{I \leq J}k_{IJ}
(n_I  m_J + n_J m_I)\theta_x\ell_y }
\end{equation}
We should recognize this as the boundary conditions for a particle on a torus with flux:
\begin{multline}
\sum_{I\leq J} k_{IJ} m_I n_J + \sum_{I\geq J} k_{IJ} m_{I}n_{J} 
= \sum_{I, J} K_{IJ} n_{I}m_{J}
\end{multline}
We see then that our Hamiltonian system has the (mixed) Hall conductance
$n\cdot K \cdot m$, in agreement with the Chern-Simons response function
derived from the spacetime path integral. In the case of $\kappa = 1$ with $n =
m = 1$, this becomes a system with integer hall conductance $K$ wich is an even
integer. 

\emph{Discussion.} We have derived a model for the $2+1$d $U^\ka(1)$ SPT states
in terms of both a spacetime lattice path integral and a commuting projector
model. We have confirmed the Hall response in two ways: by coupling the
spacetime model to a background gauge field, and by examining the ground state
wavefunction of the Hamiltonian model on a torus with twisted boundary
conditions. Now we must understand how this model evades the no-go theorem.

The infinite dimensional on-site Hilbert space of the rotors, combined with an
action that is not a continuous function of the field variables, is what allows
this model to exist. The discontinuous action is critical for commuting
projectors; if it were continuous, then one could truncate the Hilbert space to
low-angular momentum modes and render the on-site Hilbert space finite while
retaining the full $U^\ka(1)$ symmetry and commuting-projector property, hence
running afoul of the no-go theorem \cite{Kapustin:2020aa}. Conversely, the
no-go theorem assumes that the ground state wavefunction is a finite Laurent
polynomial in $e^{i\theta_x}, e^{i\theta_y}$, an assumption which is violated
in our model (See the details of the Chern number calculation in the Appendix
for an example). This commuting projector model represents a fixed-point theory for $U^\kappa(1)$ SPT phases with nonzero Hall conductance; it may be that any such fixed-point theory requires an infinite on-site Hilbert space. 

There is a connection here to the theory of discontinuous group cocycles. SPT
phases in $d+1$ dimensions are classified by $H^{d+1}(G, U(1))$
\cite{CGL1314}, and their topological field theory actions are given by
representative cocycles. Eq. (\ref{U1spt}) is an example of this sort of
cocycle. For reasons similar to the underlying argument of the no-go theorem,
if we restrict to continuous cochains, then corrsponding
$H_\text{continuous}^{2}(G, U(1))$ is not large enough to classify the group
extensions of $G$ by $U(1)$ \cite{WW1104}.   It is only by allowing
piecewise-continuous cochains, like the action in eq. (\ref{U1spt}), does
corrsponding $H^{2}(G, U(1))$ classify the group extensions, as well as the
projective representations of $G$ and the 1+1d SPT orders \cite{CGL1314}.

~

This research is partially supported by NSF DMR-2022428, the NSF Graduate Research Fellowship under Grant No. 1745302, and by the Simons
Collaboration on Ultra-Quantum Matter, which is a grant from the Simons
Foundation (651440).

\bibliography{../../bib/all,../../bib/allnew,../../bib/publst,U1SPT_Bib}


\appendix

\allowdisplaybreaks

\section{Chern Number for Hall Conductance}

Working with the wavefunction on the lattice shown in Fig. \ref{fig:HCSU}, recall that we label lattice points by $(x, y) \in [0...L_x-1]\times [0...L_y-1]$. We twist the boundary conditions by $\theta_x n_I$, $\theta_y m_I$, so that $\phi_{I, x=0, y}^\RZ= \phi_{I, x=L_x, y}^\RZ - \theta_x n_{I}$ and $\phi_{I,x, y=0}^\RZ = \phi_{I, x, y=L_y}^\RZ - \theta_y m^I$, with $n_I, m_I \in \mathbb{Z}^\kappa$.

Consider first the plaquettes marked in red. The contribution of these plaquettes to the path integral is of the form:
\begin{multline}
\exp\Big\{
-2\pi i \sum_{I \leq J}k_{IJ} 
\Big[
\\\nonumber
(\phi_{I, x=L, y}^\RZ + n_I \theta_x - \phi_{I, x=L_x - 1, y}^\RZ)
\toZ{\phi_{J, x=L_x, y+1}^\RZ - \phi_{J, x=L_x, y}^\RZ}
\\\nonumber
 -
 \Big(
(\phi_{I, x=L_x - 1, y+1}^\RZ - \phi_{I, x = L_x-1, y}^\RZ)
\\\nonumber
\times
\toZ{\phi_{J, x=L_x, y+1}^\RZ + n_J\theta_x - \phi_{J, x=L_x -1, y+1}^\RZ} 
\Big)
\Big]
\Big\}
\end{multline}
Incrementing $\theta_x$ by $\ell_x$ changes the first term by an integer, while the second term changes by $(\phi_{I, x=L_x - 1, y+1}^\RZ - \phi_{I, x = L_x-1, y}^\RZ)\ell_x$, and so the multiplicative change on the path integral is:
\begin{equation}
e^{2\pi i \sum_{I \leq J}k_{IJ} (\phi_{I, x=L_x - 1, y+1}^\RZ - \phi_{I, x = L_x-1, y}^\RZ)n_J\ell_x }
\end{equation}
Similarly, the plaquettes marked in blue change by:
\begin{equation}
e^{-2\pi i \sum_{I \leq J}k_{IJ} (\phi_{I, x+1, y=L_y-1}^\RZ - \phi_{I, x, y=L_y-1}^\RZ)m_J\ell_y }
\end{equation}
On the purple plaquettes at the corner, the contribution to the path integral is:
\begin{multline}
\exp\Big\{
-2\pi i \sum_{I \leq J}k_{IJ} 
\Big[
\\\nonumber
\Big((\phi_{I, x=L_x, y=L_y-1}^\RZ + n_I \theta_x - \phi_{I,x=L_x-, y=L_y-1}^\RZ)
\\\nonumber
\times\toZ{\phi_{J, x=L_x, y=L_y}^\RZ + m_J\theta_y - \phi_{J, x=L_x, y=L_y-1}^\RZ} 
\Big)
\\\nonumber
-\Big(
(\phi_{I, x=L_x-1, y=L_y}^\RZ + m_I\theta_y - \phi_{I, x=L_x-1, y=L_y-1}^\RZ)
\\\nonumber
\times\toZ{\phi_{J, x=L_x, y=L_y}^\RZ + n_J\theta_x - \phi_{J, x=L_x-1, y=L_y}^\RZ} 
\Big)
\Big]
\Big\}
\end{multline}
which will change by:
\begin{multline}
\exp\Big\{
-2\pi i \sum_{I \leq J}k_{IJ} 
\Big[
\\
(\phi_{I, x=L_x, y=L_y-1}^\RZ + n_I \theta_x - \phi_{I,x=L_x-, y=L_y-1}^\RZ) m_J\ell_y
\\
-(\phi_{I, x=L_x-1, y=L_y}^\RZ + m_I\theta_y - \phi_{I, x=L_x-1, y=L_y-1}^\RZ)n_J\ell_x 
\Big]
\Big\}
\end{multline}
Combining the change on the red, blue, and purple plaquettes, the overall change to the wavefunction is:
\begin{equation}
e^{-2\pi i \sum_{I \leq J}k_{IJ} \left[
n_I \theta_x m_J\ell_y
-m_I\theta_y n_J\ell_x
- n_{J}\ell_x \int_{\gamma_1}d\phi_{I}^\RZ 
+ m_{J}\ell_Y\int_{\gamma_2} d\phi_{I}^\RZ
\right] }
\end{equation}
where $\gamma_1, \gamma_2$ are the red and blue loops in Fig. \ref{fig:HCSU}, respectively. As that $\gamma_1, \gamma_2$ are closed, the sums along those curves vanish, and we are left with the result in the main text.

\section{Reducing to $\mathbb Z_n$ gauge theory}

When $\ka=1$, the above becomes
\begin{align}
 \label{U1spt1}
 Z &=\int [\prod\dd \phi^{\RZ}]\ 
 \ee^{\ii 2\pi k \int_{\cM^3} 
\dd \phi^\RZ \dd \toZ{\dd \phi^\RZ} 
},
\end{align}
which describe a $U(1)$ SPT state with Hall conductance
$\si_{xy}=\frac{2k}{2\pi}$.

Let us compare the model \eq{U1spt1} for $U(1)$ SPT state with a model for
$\Z_n$ SPT state \cite{PhysRevB.95.205142}:
\begin{align}
 \label{Znspt}
 Z &=\sum_{\phi^{\Z_n}} 
 \ee^{\ii \frac{2\pi k}{n^2} \int_{\cM^3} 
a^{\Z_n} \dd a^{\Z_n}
},
\nonumber\\
a^{\Z_n} &= \dd \phi^{\Z_n} - n\toZ{\frac1n \dd \phi^{\Z_n}},
\end{align}
where $\phi^{\Z_n}$ is a $\Z_n$-valued 0-cochain.  The above can be rewritten
as
\begin{align}
 \label{Znspt1}
 Z &=\sum_{\phi^{\Z_n}} 
 \ee^{-\ii 2\pi k \int_{\cM^3} 
\dd \phi^\RZ \dd \toZ{\dd \phi^\RZ}
},
&
\phi^\RZ &=\frac1n \phi^{\Z_n}
.
\end{align}
We see that the model for the $\Z_n$ SPT state and the model for the $U(1)$ SPT
state have very similar forms.  Here, $\dd \phi^\RZ \dd \toZ{\dd \phi^\RZ}$
with $ \phi^\RZ =\frac1n \phi^{\Z_n}$ is a cocycle in $H^3(\Z_n, \RZ)$, while
$\dd \phi^\RZ \dd \toZ{\dd \phi^\RZ}$ is a cocycle in $H^3(U(1), \RZ)$.

\section{Properties of the Commuting Projectors}

Recall the definition of the $\hat M_i$:
\begin{multline}
M_{\phi_{I, 4}^\RZ \to \phi_{I, 5}^\RZ}(\phi_{I, 1}^\RZ, ..., \phi_{I, 8}^\RZ)
=
\\ 
\exp\Big\{
2 \pi i \sum_{I\leq J}k_{IJ}\Big(
  \phi_{I,0}^\RZ(\toZ{\phi_{J,5}^\RZ - \phi_{J,2}^\RZ} + \toZ{\phi_{J,3}^\RZ - \phi_{J,5}^\RZ} \\  + \toZ{\phi_{J,4}^\RZ-\phi_{J,3}^\RZ} + \toZ{\phi_{J,2}^\RZ - \phi_{J,4}^\RZ})\\ 
+ \phi_{I,2}^\RZ(\toZ{\phi_{J,5}^\RZ - \phi_{J,6}^\RZ} + \toZ{\phi_{J,2}^\RZ - \phi_{J,5}^\RZ} \\  + \toZ{\phi_{J,4}^\RZ - \phi_{J,2}^\RZ} + \toZ{\phi_{J,6}^\RZ - \phi_{J,4}^\RZ})\\ 
+ \phi_{I,3}^\RZ(\toZ{\phi_{J,7}^\RZ - \phi_{J,5}^\RZ} + \toZ{\phi_{J,4}^\RZ - \phi_{J,7}^\RZ} \\  + \toZ{\phi_{J,3}^\RZ - \phi_{J,4}^\RZ} + \toZ{\phi_{J,5}^\RZ - \phi_{J,3}^\RZ})\\ 
+ \phi_{I,5}^\RZ(\toZ{\phi_{J,6}^\RZ - \phi_{J,5}^\RZ} + \toZ{\phi_{J,8}^\RZ - \phi_{J,6}^\RZ} \\  + \toZ{\phi_{J,7}^\RZ - \phi_{J,8}^\RZ} + \toZ{\phi_{J,5}^\RZ - \phi_{J,7}^\RZ})\\ 
+ \phi_{I,4}^\RZ(\toZ{\phi_{J,7}^\RZ - \phi_{J,4}^\RZ} + \toZ{\phi_{J,8}^\RZ - \phi_{J,7}^\RZ} \\  + \toZ{\phi_{J,6}^\RZ - \phi_{J,8}^\RZ} + \toZ{\phi_{J,4}^\RZ - \phi_{J,6}^\RZ})
\Big)
\Big\}\label{eq:M_itojRepeat}
\end{multline}
Here we walk through the calculations to show analytically that the $\hat M_i$ are hermitian, $U^\ka(1)$ symmetric, commuting projectors.

For hermiticity, we wish to show that
\begin{equation}
M_{\phi_{I, 5}^\RZ \to \phi_{I, 4}^\RZ}(\phi_{I, 1}^\RZ, ..., \phi_{I, 8}^\RZ)^*
=
M_{\phi_{I, 4}^\RZ \to \phi_{I, 5}^\RZ}(\phi_{I, 1}^\RZ, ..., \phi_{I, 8}^\RZ)
\end{equation}
To see this, first note that the terms in eq. (\ref{eq:M_itojRepeat}) with coefficients of $\phi_{I, 0}^\RZ, \phi_{I, 2}^\RZ$, or $\phi_{I, 3}^\RZ$ are antisymmetric under $\phi_{I, 4}^\RZ \leftrightarrow \phi_{I, 5}^\RZ$. Next, consider the $\phi_{I, 5}^\RZ$ term:
\begin{multline}
\phi_{I,5}^\RZ(\toZ{\phi_{J,6}^\RZ - \phi_{J,5}^\RZ} + \toZ{\phi_{J,8}^\RZ - \phi_{J,6}^\RZ} \\  + \toZ{\phi_{J,7}^\RZ - \phi_{J,8}^\RZ} + \toZ{\phi_{J,5}^\RZ - \phi_{J,7}^\RZ})
\end{multline}
under $\phi_{I, 4}^\RZ \leftrightarrow \phi_{I, 5}^\RZ$, this becomes:
\begin{multline}
-\phi_{I,4}^\RZ(\toZ{\phi_{J,4}^\RZ - \phi_{J,6}^\RZ} + \toZ{\phi_{J,6}^\RZ - \phi_{J,8}^\RZ} \\  + \toZ{\phi_{J,8}^\RZ - \phi_{J,7}^\RZ} + \toZ{\phi_{J,7}^\RZ - \phi_{J,4}^\RZ})
\end{multline}
Which is precisely $-1$ times the $\phi_{I, 4}^\RZ$ term. Similarly, the $\phi_{I, 4}^\RZ$ term becomes minus the $\phi_{I, 5}^\RZ$ term. Taking all of these together with the minus sign from the factor of $i$, we see that the $\hat M_i$ are hermitian.

The symmetry $\phi_{I}^\RZ \to \phi_{I}^\RZ + \theta_{I}$ arises essentially because only $d\phi_I^\RZ$ appears in action $2\pi \sum_{I\leq J} k_{IJ} d\phi_{I}^\RZ\toZ{d\phi_J^\RZ}$. To see this explicitly in eq. \ref{eq:M_itojRepeat}, first note that the $\theta_I$ cancel in the rounded terms. What remains is:
\begin{widetext}
\begin{multline}
M_{\phi_{I, 4}^\RZ \to \phi_{I, 5}^\RZ}(\phi_{I, 1}^\RZ, ..., \phi_{I, 8}^\RZ)
=
\\ 
\exp\Big\{
2 \pi i \sum_{I\leq J}k_{IJ}\Big(
  (\phi_{I,0}^\RZ + \theta_I)(\toZ{\phi_{J,5}^\RZ - \phi_{J,2}^\RZ} + \toZ{\phi_{J,3}^\RZ - \phi_{J,5}^\RZ}   + \toZ{\phi_{J,4}^\RZ-\phi_{J,3}^\RZ} + \toZ{\phi_{J,2}^\RZ - \phi_{J,4}^\RZ})\\ 
+ (\phi_{I,2}^\RZ + \theta_I)(\toZ{\phi_{J,5}^\RZ - \phi_{J,6}^\RZ} + \toZ{\phi_{J,2}^\RZ - \phi_{J,5}^\RZ}  + \toZ{\phi_{J,4}^\RZ - \phi_{J,2}^\RZ} + \toZ{\phi_{J,6}^\RZ - \phi_{J,4}^\RZ})\\ 
+ (\phi_{I,3}^\RZ + \theta_I)(\toZ{\phi_{J,7}^\RZ - \phi_{J,5}^\RZ} + \toZ{\phi_{J,4}^\RZ - \phi_{J,7}^\RZ} + \toZ{\phi_{J,3}^\RZ - \phi_{J,4}^\RZ} + \toZ{\phi_{J,5}^\RZ - \phi_{J,3}^\RZ})\\ 
+ (\phi_{I,5}^\RZ + \theta_I)(\toZ{\phi_{J,6}^\RZ - \phi_{J,5}^\RZ} + \toZ{\phi_{J,8}^\RZ - \phi_{J,6}^\RZ} + \toZ{\phi_{J,7}^\RZ - \phi_{J,8}^\RZ} + \toZ{\phi_{J,5}^\RZ - \phi_{J,7}^\RZ})\\ 
+ (\phi_{I,4}^\RZ + \theta_I))(\toZ{\phi_{J,7}^\RZ - \phi_{J,4}^\RZ} + \toZ{\phi_{J,8}^\RZ - \phi_{J,7}^\RZ} + \toZ{\phi_{J,6}^\RZ - \phi_{J,8}^\RZ} + \toZ{\phi_{J,4}^\RZ - \phi_{J,6}^\RZ})
\Big)
\\
=
M_{\phi_{I, 4}^\RZ \to \phi_{I, 5}^\RZ}(\phi_{I, 1}^\RZ, ..., \phi_{I, 8}^\RZ)
\exp\Big\{
2 \pi i \sum_{I\leq J}k_{IJ}
\theta_I\Big(\toZ{\phi_{J,5}^\RZ - \phi_{J,2}^\RZ} + \toZ{\phi_{J,3}^\RZ - \phi_{J,5}^\RZ} \\  + \toZ{\phi_{J,4}^\RZ-\phi_{J,3}^\RZ} + \toZ{\phi_{J,2}^\RZ - \phi_{J,4}^\RZ}
+\toZ{\phi_{J,5}^\RZ - \phi_{J,6}^\RZ} + \toZ{\phi_{J,2}^\RZ - \phi_{J,5}^\RZ}   + \toZ{\phi_{J,4}^\RZ - \phi_{J,2}^\RZ} + \toZ{\phi_{J,6}^\RZ - \phi_{J,4}^\RZ}
+\toZ{\phi_{J,7}^\RZ - \phi_{J,5}^\RZ} + \toZ{\phi_{J,4}^\RZ - \phi_{J,7}^\RZ} \\  + \toZ{\phi_{J,3}^\RZ - \phi_{J,4}^\RZ} + \toZ{\phi_{J,5}^\RZ - \phi_{J,3}^\RZ}
+\toZ{\phi_{J,6}^\RZ - \phi_{J,5}^\RZ} + \toZ{\phi_{J,8}^\RZ - \phi_{J,6}^\RZ} + \toZ{\phi_{J,7}^\RZ - \phi_{J,8}^\RZ} + \toZ{\phi_{J,5}^\RZ - \phi_{J,7}^\RZ}
+\toZ{\phi_{J,7}^\RZ - \phi_{J,4}^\RZ} + \toZ{\phi_{J,8}^\RZ - \phi_{J,7}^\RZ} \\  + \toZ{\phi_{J,6}^\RZ - \phi_{J,8}^\RZ} + \toZ{\phi_{J,4}^\RZ - \phi_{J,6}^\RZ}
\Big)\Big\}
=
M_{\phi_{I, 4}^\RZ \to \phi_{I, 5}^\RZ}(\phi_{I, 1}^\RZ, ..., \phi_{I, 8}^\RZ)
\end{multline}
\end{widetext}
One may check that remaining rounded terms cancel one-by-one, essentially because this is $\dd \toZ{\dd \phi_{I}^\RZ}$ evaluated over the closed surface of the complex. To see that the $\hat M_i$ are symmetric under $\phi_{I, i}^\RZ \to \phi_{I, i}^\RZ + n_{I, i}$, first note that, because $k_{IJ}$ is integral, we need only to worry about the rounded terms. However, one may check that each the effect $n_{I, i}$ in each sum of rounded terms cancels, essentially because each sum is $\dd\toZ{\dd\phi_I^\RZ}$, and under $\phi_{I}^\RZ \to \phi_{I}^\RZ + n_{I}$ this becomes $\dd\toZ{\dd\phi_I^\RZ +\dd n_I} = \dd\toZ{\dd\phi_I^\RZ} + \dd^2 n_I$ and $\dd^2 =0$. All told, we now see that the $\hat M_i$ are symmetric under $\phi_{I, i}^\RZ \to \phi_{I, i}^\RZ + n_{I, i} + \theta_I$, i.e. the $\hat M_i$ are $U^\ka(1)$ symmetric. 

\begin{figure}
\includegraphics[width = .5\columnwidth]{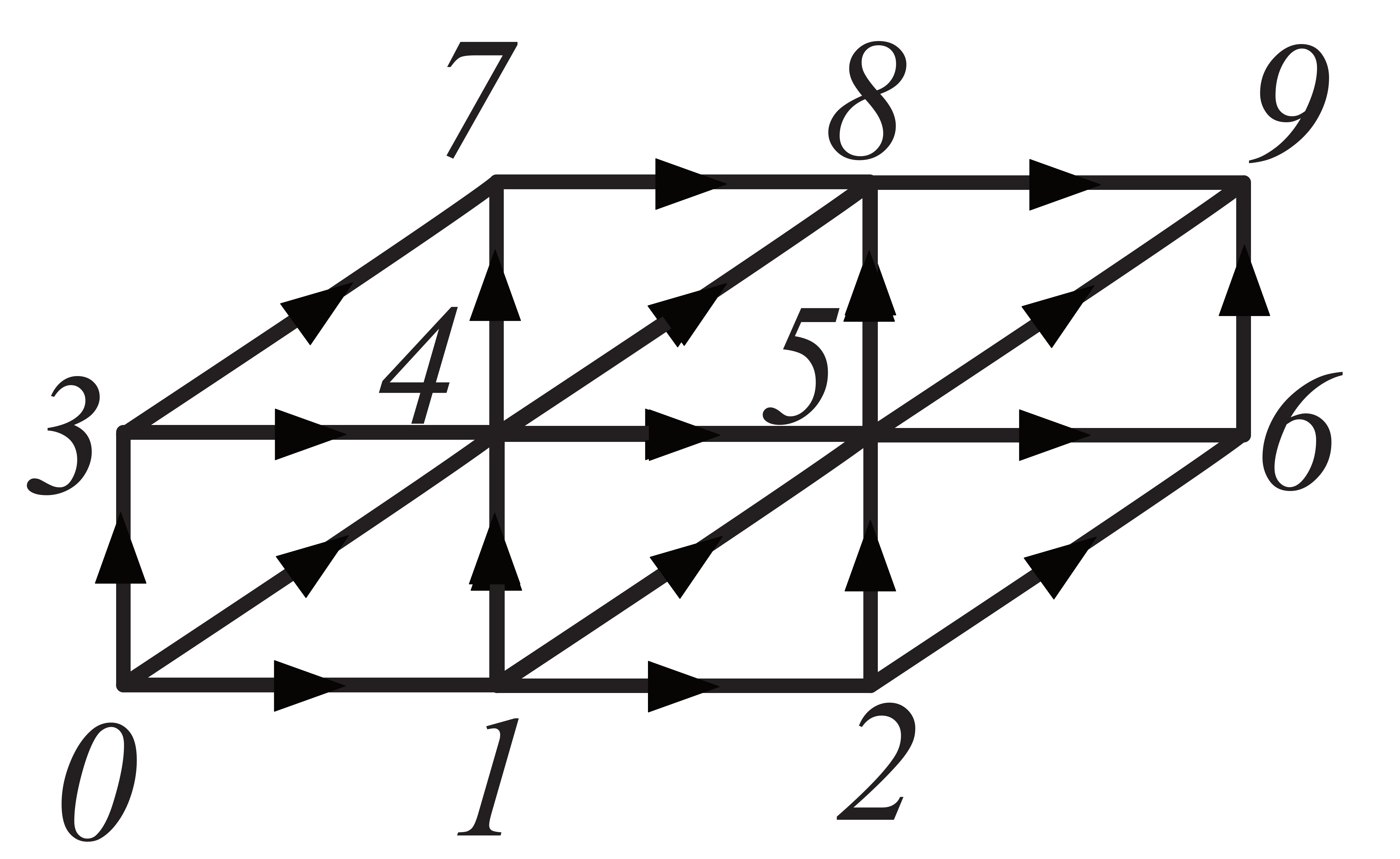}
\caption{}
\label{fig:CommutingSetUp}
\end{figure}

Next we check commutation. The only nontrivial case occurs when the $\hat M_i$ are on adjacent sites. For this calculation, we will use a slightly different convention, indicating the time-evolved points with a prime as opposed to a new number, so that amplitudes take the form $M_{\phi_{I, 4}^\RZ \to (\phi_{I, 4}^\RZ)'}(...)$. We also drop the $\RZ$ superscripts. Consider then the $2d$ spatial complex in Fig. \ref{fig:CommutingSetUp}. We wish to compare:
\begin{align}
\hat M_4 \hat M_5 = M_{\phi_{I, 4} \to \phi_{I, 4}'}(\phi_{I, 1}, ..., \phi_{I, 5}',..., \phi_{I, 8})
\\ \nonumber
\times M_{\phi_{I, 5} \to \phi_{I, 5}'}(\phi_{I, 1}, ..., \phi_{I, 8})
\label{eq:Commuting1}
\end{align}
to
\begin{align}
\hat M_5 \hat M_4 = M_{\phi_{I, 5} \to \phi_{I, 5}'}(\phi_{I, 1}, ..., \phi_{I, 4}',..., \phi_{I, 8})
\\ \nonumber
\times M_{\phi_{I, 4} \to \phi_{I, 4}'}(\phi_{I, 1}, ..., \phi_{I, 8})
\end{align}
Expanding eq. (\ref{eq:Commuting1}), it becomes:
\begin{widetext}
\begin{multline}
\hat M_5 \hat M_4  = \exp\Bigg\{
2\pi i \sum_{I\leq J}k_{IJ}
\Bigg(
\phi_{I, 0}(
\toZ{\phi_{J, 4}' - \phi_{J, 1}}
+ \toZ{\phi_{J, 3} - \phi_{J, 4}'}
+ \toZ{\phi_{J, 4} - \phi_{J, 3}}
+ \toZ{\phi_{J, 1} - \phi_{J, 4}})
\\
+ \phi_{I, 1}(
\toZ{\phi_{J, 4}' - \phi_{J, 5}}
+ \toZ{\phi_{J, 1} - \phi_{J, 4}'}
+ \toZ{\phi_{J, 4} - \phi_{J, 1}}
+ \toZ{\phi_{J,  5} - \phi_{J, 4}})
+ \phi_{I, 3}(
\toZ{\phi_{J, 4}' - \phi_{J, 3}}
+ \toZ{\phi_{J, 7} - \phi_{J, 4}'}
+ \toZ{\phi_{J, 4} - \phi_{J, 7}}
+ \toZ{\phi_{J, 3} - \phi_{J, 4}})
\\
+ \phi_{I, 4}(
\toZ{\phi_{J, 7} - \phi_{J, 4}}
+ \toZ{\phi_{J, 8} - \phi_{J, 7}}
+ \toZ{\phi_{J, 5} - \phi_{J, 8}}
+ \toZ{\phi_{J, 4} - \phi_{J, 5}})
+ \phi_{I, 4}'(
\toZ{\phi_{J, 5} - \phi_{J, 4}'}
+ \toZ{\phi_{J, 8} - \phi_{J, 5}}
+ \toZ{\phi_{J, 7} - \phi_{J, 8}}
+ \toZ{\phi_{J, 4}' - \phi_{J, 7}})
\\
+\phi_{I, 1} (
\toZ{\phi_{J, 5}' - \phi_{J, 2}}
+ \toZ{\phi_{J, 4}' - \phi_{J, 5}'}
+ \toZ{\phi_{J, 5} - \phi_{J, 4}'}
+ \toZ{\phi_{J, 2} - \phi_{J, 5}})
+ \phi_{I, 2}(
\toZ{\phi_{J, 5}' - \phi_{J, 6}}
+ \toZ{\phi_{J, 2} - \phi_{J, 5}'}
+ \toZ{\phi_{J, 5} - \phi_{J, 2}}
+ \toZ{\phi_{J, 6} - \phi_{J, 5}})
\\
+\phi_{I, 4}'(
\toZ{\phi_{J, 5}' - \phi_{J, 4}'}
+\toZ{\phi_{J, 8} - \phi_{J, 5}'}
+\toZ{\phi_{J, 5} - \phi_{J, 8}}
+\toZ{\phi_{J, 4}' - \phi_{J, 5}})
+\phi_{I, 5}(
\toZ{\phi_{J, 8} - \phi_{J, 5}}
+ \toZ{\phi_{J, 9} - \phi_{J, 8}}
+ \toZ{\phi_{J, 6} - \phi_{J, 9}}
+ \toZ{\phi_{J, 5} - \phi_{J, 6}}
)
\\
+ \phi_{I, 5}'(
\toZ{\phi_{J, 6} - \phi_{J, 5}'}
+ \toZ{\phi_{J, 9} - \phi_{J, 6}}
+ \toZ{\phi_{J, 8} - \phi_{J, 9}}
+ \toZ{\phi_{J, 5}' - \phi_{J, 8}})
\Bigg)
\Bigg\}
\end{multline}
On the other hand,
\begin{multline}
\hat M_4 \hat M_5  
=
\exp\Bigg\{
2\pi i \sum_{I\leq J}k_{IJ}
\Bigg(
\phi_{I, 0}(
\toZ{\phi_{J, 4}' - \phi_{J, 1}}
+ \toZ{\phi_{J, 3} - \phi_{J, 4}'}
+ \toZ{\phi_{J, 4} - \phi_{J, 3}}
+ \toZ{\phi_{J, 1} - \phi_{J, 4}})
\\
+ \phi_{I, 1}(
\toZ{\phi_{J, 4}' - \phi_{J, 5}'}
+ \toZ{\phi_{J, 1} - \phi_{J, 4}'}
+ \toZ{\phi_{J, 4} - \phi_{J, 1}}
+ \toZ{\phi_{J,  5}' - \phi_{J, 4}})
+ \phi_{I, 3}(
\toZ{\phi_{J, 4}' - \phi_{J, 3}}
+ \toZ{\phi_{J, 7} - \phi_{J, 4}'}
+ \toZ{\phi_{J, 4} - \phi_{J, 7}}
+ \toZ{\phi_{J, 3} - \phi_{J, 4}})
\\
+ \phi_{I, 4}(
\toZ{\phi_{J, 7} - \phi_{J, 4}}
+ \toZ{\phi_{J, 8} - \phi_{J, 7}}
+ \toZ{\phi_{J, 5}' - \phi_{J, 8}}
+ \toZ{\phi_{J, 4} - \phi_{J, 5}'})
+ \phi_{I, 4}'(
\toZ{\phi_{J, 5}' - \phi_{J, 4}'}
+ \toZ{\phi_{J, 8} - \phi_{J, 5}'}
+ \toZ{\phi_{J, 7} - \phi_{J, 8}}
+ \toZ{\phi_{J, 4}' - \phi_{J, 7}})
\\
+\phi_{I, 1} (
\toZ{\phi_{J, 5}' - \phi_{J, 2}}
+ \toZ{\phi_{J, 4} - \phi_{J, 5}'}
+ \toZ{\phi_{J, 5} - \phi_{J, 4}}
+ \toZ{\phi_{J, 2} - \phi_{J, 5}})
+ \phi_{I, 2}(
\toZ{\phi_{J, 5}' - \phi_{J, 6}}
+ \toZ{\phi_{J, 2} - \phi_{J, 5}'}
+ \toZ{\phi_{J, 5} - \phi_{J, 2}}
+ \toZ{\phi_{J, 6} - \phi_{J, 5}})
\\
+\phi_{I, 4}(
\toZ{\phi_{J, 5}' - \phi_{J, 4}}
+\toZ{\phi_{J, 8} - \phi_{J, 5}'}
+\toZ{\phi_{J, 5} - \phi_{J, 8}}
+\toZ{\phi_{J, 4} - \phi_{J, 5}})
+\phi_{I, 5}(
\toZ{\phi_{J, 8} - \phi_{J, 5}}
+ \toZ{\phi_{J, 9} - \phi_{J, 8}}
+ \toZ{\phi_{J, 6} - \phi_{J, 9}}
+ \toZ{\phi_{J, 5} - \phi_{J, 6}}
)
\\
+ \phi_{I, 5}'(
\toZ{\phi_{J, 6} - \phi_{J, 5}'}
+ \toZ{\phi_{J, 9} - \phi_{J, 6}}
+ \toZ{\phi_{J, 8} - \phi_{J, 9}}
+ \toZ{\phi_{J, 5}' - \phi_{J, 8}})
\Bigg)
\Bigg\}
\end{multline}
\end{widetext} 
Proceeding term-by-term, one may verify that these are equal. (To simplify the calculation, note that the terms with coefficients $\phi_{I, 0}, \phi_{I, 2}, \phi_{I, 3}, \phi_{I, 5}$, and $\phi_{I, 5}'$ are identical.)

We also verify that the $\hat M_i$ are projectors. Retaining the same notation, we calculate:
\begin{widetext}
\begin{multline}
M_{\phi_{I, 4}' \to \phi_{I, 4}''}(\phi_{I, 1}, ..., \phi_{I, 4}', ..., \phi_{I, 8})
M_{\phi_{I, 4} \to \phi_{I, 4}'}(\phi_{I, 1}, ..., \phi_{I, 8})
\\
=
\exp\Bigg\{
2\pi i \sum_{I\leq J}k_{IJ}
\Bigg(
\phi_{I, 0}(
\toZ{\phi_{J, 4}' - \phi_{J, 1}}
+ \toZ{\phi_{J, 3} - \phi_{J, 4}'}
+ \toZ{\phi_{J, 4} - \phi_{J, 3}}
+ \toZ{\phi_{J, 1} - \phi_{J, 4}})
\\
+ \phi_{I, 1}(
\toZ{\phi_{J, 4}' - \phi_{J, 5}}
+ \toZ{\phi_{J, 1} - \phi_{J, 4}'}
+ \toZ{\phi_{J, 4} - \phi_{J, 1}}
+ \toZ{\phi_{J,  5} - \phi_{J, 4}})
+ \phi_{I, 3}(
\toZ{\phi_{J, 4}' - \phi_{J, 3}}
+ \toZ{\phi_{J, 7} - \phi_{J, 4}'}
+ \toZ{\phi_{J, 4} - \phi_{J, 7}}
+ \toZ{\phi_{J, 3} - \phi_{J, 4}})
\\
+ \phi_{I, 4}(
\toZ{\phi_{J, 7} - \phi_{J, 4}}
+ \toZ{\phi_{J, 8} - \phi_{J, 7}}
+ \toZ{\phi_{J, 5} - \phi_{J, 8}}
+ \toZ{\phi_{J, 4} - \phi_{J, 5}})
+ \phi_{I, 4}'(
\toZ{\phi_{J, 5} - \phi_{J, 4}'}
+ \toZ{\phi_{J, 8} - \phi_{J, 5}}
+ \toZ{\phi_{J, 7} - \phi_{J, 8}}
+ \toZ{\phi_{J, 4}' - \phi_{J, 7}})
\\
+\phi_{I, 0}(
\toZ{\phi_{J, 4}'' - \phi_{J, 1}}
+ \toZ{\phi_{J, 3} - \phi_{J, 4}''}
+ \toZ{\phi_{J, 4}' - \phi_{J, 3}}
+ \toZ{\phi_{J, 1} - \phi_{J, 4}'})
+ \phi_{I, 1}(
\toZ{\phi_{J, 4}'' - \phi_{J, 5}}
+ \toZ{\phi_{J, 1} - \phi_{J, 4}''}
+ \toZ{\phi_{J, 4}' - \phi_{J, 1}}
+ \toZ{\phi_{J,  5} - \phi_{J, 4}'})
\\
+ \phi_{I, 3}(
\toZ{\phi_{J, 4}'' - \phi_{J, 3}}
+ \toZ{\phi_{J, 7} - \phi_{J, 4}''}
+ \toZ{\phi_{J, 4}' - \phi_{J, 7}}
+ \toZ{\phi_{J, 3} - \phi_{J, 4}'})
+ \phi_{I, 4}'(
\toZ{\phi_{J, 7} - \phi_{J, 4}'}
+ \toZ{\phi_{J, 8} - \phi_{J, 7}}
+ \toZ{\phi_{J, 5} - \phi_{J, 8}}
+ \toZ{\phi_{J, 4}' - \phi_{J, 5}})
\\
+ \phi_{I, 4}''(
\toZ{\phi_{J, 5} - \phi_{J, 4}''}
+ \toZ{\phi_{J, 8} - \phi_{J, 5}}
+ \toZ{\phi_{J, 7} - \phi_{J, 8}}
+ \toZ{\phi_{J, 4}'' - \phi_{J, 7}})
\Bigg)
\Bigg\}
\\
=
\exp\Bigg\{
2\pi i \sum_{I\leq J}k_{IJ}
\Bigg(
\phi_{I, 0}(
\toZ{\phi_{J, 4}'' - \phi_{J, 1}}
+ \toZ{\phi_{J, 3} - \phi_{J, 4}''}
+ \toZ{\phi_{J, 4} - \phi_{J, 3}}
+ \toZ{\phi_{J, 1} - \phi_{J, 4}})
\\
+ \phi_{I, 1}(
\toZ{\phi_{J, 4}'' - \phi_{J, 5}}
+ \toZ{\phi_{J, 1} - \phi_{J, 4}''}
+ \toZ{\phi_{J, 4} - \phi_{J, 1}}
+ \toZ{\phi_{J,  5} - \phi_{J, 4}})
+ \phi_{I, 3}(
\toZ{\phi_{J, 4}'' - \phi_{J, 3}}
+ \toZ{\phi_{J, 7} - \phi_{J, 4}''}
+ \toZ{\phi_{J, 4} - \phi_{J, 7}}
+ \toZ{\phi_{J, 3} - \phi_{J, 4}})
\\
+ \phi_{I, 4}(
\toZ{\phi_{J, 7} - \phi_{J, 4}}
+ \toZ{\phi_{J, 8} - \phi_{J, 7}}
+ \toZ{\phi_{J, 5} - \phi_{J, 8}}
+ \toZ{\phi_{J, 4} - \phi_{J, 5}})
+ \phi_{I, 4}''(
\toZ{\phi_{J, 5} - \phi_{J, 4}''}
+ \toZ{\phi_{J, 8} - \phi_{J, 5}}
+ \toZ{\phi_{J, 7} - \phi_{J, 8}}
+ \toZ{\phi_{J, 4}'' - \phi_{J, 7}})
\Bigg)
\Bigg\}
\\
= M_{\phi_{I, 4} \to \phi_{I, 4}''}(\phi_{I, 1}, ..., \phi_{I, 8})
\end{multline}
\end{widetext}
Whence $\hat M_{i=4}^2 = \hat M_{i=4}$, and by translation $\hat M_i^2 = \hat M_i$ for the entire lattice.

Finally, we also note that the $\hat M_i$ are mutually independent, so there is no condition which could allow extra ground state degeneracy as in the toric code.

\section{Rewriting of the $\hat M_i$}
Defining
\begin{multline}
f(\phi_*)
=
\sum_{I\leq J} k_{IJ} \Big[\phi_{I, *}(\toZ{\phi_{J, 6}^\RZ - \phi_{J, 5}^\RZ}
+ \toZ{\phi_{J, 8} - \phi_{J, 6}} \\
+ \toZ{\phi_{J, 7} - \phi_{J, 8}} + \toZ{\phi_{J, 5} - \phi_{J, 7}})\\
+ \phi_{I, 0}(\toZ{\phi_{J,5} - \phi_{J,2}} + \toZ{\phi_{J, 3} - \phi_{J,5}})\\
+ \phi_{I, 2}(\toZ{\phi_{J, 5} - \phi_{J,6}} + \toZ{\phi_{J,2} - \phi_{J,5}})\\
+ \phi_{I, 3}(\toZ{\phi_{J,7} - \phi_{J,5}} + \toZ{\phi_{J,5} - \phi_{J,3}})\Big]
\end{multline}
we can see that eq. (\ref{eq:M_itojRepeat}) may be rewritten as:
\begin{multline}
M_{\phi_{I, 4}^\RZ \to \phi_{I, 5}^\RZ}(...)(\phi_{I, 5}^\RZ)
\\
=
\exp(2\pi i f(\phi_{I, 5}^\RZ))
\exp(-2\pi i f(\phi_{I, 4}^\RZ)
\end{multline}

\end{document}